\documentclass[%
 reprint,
 amsmath,amssymb,
 aps,
]{revtex4-1}

\usepackage{graphicx}
\usepackage{dcolumn}
\usepackage{bm}
\usepackage{hyperref}

\newcommand{\erfc}{\text{erfc}}
\begin{document}

\preprint{APS/123-QED}

\title{Detailed Characterization of Rough Surfaces for Silica Materials}
\author{Timur Aslyamov}
\email{t.aslyamov@skoltech.ru}
\affiliation{Center for Design, Manufacturing and Materials, Skolkovo Institute of Science and Technology,
  Bolshoy Boulevard 30, bld. 1, Moscow, Russia 121205
}
\author{Aleksey Khlyupin}
\affiliation{Moscow Institute of Physics and Technology, Institutskiy Pereulok 9, Dolgoprudny, Moscow 141700, Russia}
\author{Vera Pletneva}
\affiliation{Schlumberger Moscow Research Center, 13 Pudovkina str., Moscow 119285, Russia}
\author{Iskander Akhatov}
\affiliation{Center for Design, Manufacturing and Materials,
   Skolkovo Institute of Science and Technology,
  Bolshoy Boulevard 30, bld. 1, Moscow, Russia 121205}

\date{\today}

\begin{abstract}
We propose a new approach to obtain the nanoscale morphology of rough surfaces from low-temperature adsorption experiments. Our method is based on one of the most realistic models of rough surfaces formulated in terms of random correlated processes and random surface density functional theory (RS-DFT) as a theoretical adsorption model. We consider the roughness in the normal direction, the correlation length of the lateral surface structure and the specific surface area as tuning parameters of RS-DFT to fit the experimental data in the low pressure range, where the influence of the surface geometry is the most crucial. One of the major advantages of the proposed approach over published methods is the best-fit detailed geometry of rough surfaces, which provides full information for further atomistic modeling. The obtained geometry correctly reflects how the nanoroughness of silica materials depends on the synthesis conditions. We demonstrate that the surface fractal dimension observed in many experiments is natural for the correlated random surface model. We investigated the surface geometry of popular silica materials synthesized at different conditions. The obtained roughness parameters and fractal dimensions coincide well with the published experimental data. Analysis of the best fit specific surface area reveals the mechanism of adsorption on rough surfaces and provides a new strategy for the search of optimal storage materials. 
\end{abstract}


\maketitle

Adsorption properties are crucial in a wide variety of applications, including CO$_2$ capture, CH$_4$ storage, and heterogeneous catalysis \cite{ferey2011hybrid, hartmann2005ordered}. Most materials of technological interest demonstrate nanoscale surface roughness. A characteristic example is silica materials with a wide range of pore sizes depending on the synthesis conditions; specifically, mesoporous materials and their modifications are used in gas storage technologies \cite{wei2018fabrication, vilarrasa2014co2}.o
Additionally, high chemical and mechanical stability make silica materials widely used as inorganic supports for affinity chromatography. To immobilize enzymes, controlled pore glass (CPG) materials \cite{weetall1993preparation} and silica gels \cite{avnir1994enzymes} are used. Additionally, experimental investigations of silica glass using small angle X-ray scattering (SAXS) have claimed the presence of a roughness region $2\delta$ around $15-20$ \r{A} \cite{levitz1991porous, mitropoulos1995characterization,mitropoulos2015formation}. Almost all popular silica materials have been investigated using fractal approaches, where the deviation of the fractal dimensions from the flat plane $D_f-2$ indicates significant roughness \cite{sonwane1999characterization, smith2010fractal, coasne2012atomistic}.

\begin{figure}[b]
\includegraphics[width=8cm]{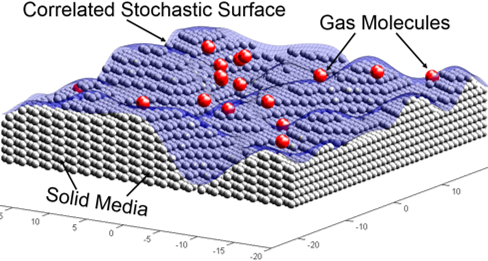}
\caption{\label{fig:rough_surface} An example of the rough surfaces described in this work. The surface (blue colour) is a realization of the correlated random process, where the variance defines the roughness in the normal direction and the correlation length is the scale of the lateral structure. 
}
\end{figure}

Several groups of researchers have demonstrated that adsorption properties for rough solid substrates can be described by theoretical models accounting for the surface geometry \cite{ravikovitch2006density, neimark2009quenched, jagiello2013carbon, jagiello2015dual, aslyamov2017density, aslyamov2019random}. Therefore, an inverse problem can be formulated to obtain the surface geometry from well-defined low-temperature adsorption measurements. The direct correlation between adsorption properties and detailed surface geometry can be used as insight for the surface modification process to design the optimal storage \cite{serna2010influence, chen2017amine} or catalysis materials \cite{galarneau2009optimization, smith2014surface}. However, the majority of rough surface approaches use oversimplified geometrical models that allow adsorption experiments to be fit but do not provide a correct description of the realistic roughness.
In our study, we characterized the nanoscale morphology of popular silica materials, accounting for the properties of realistic rough surfaces. Our obtained results fit published experimental data well and correctly reflect the material synthesis conditions. We have also demonstrated that the fractal dimension observed in experiments is derived from a correlated random surface model. Thus, we have provided a very detailed characterization of nanorough surfaces in comparison with published alternative theoretical approaches. The obtained surface characteristics allow to reconstruct an explicit molecular model, which can be used as input data for further atomistic simulations \cite{shi2019bottom, carr2011microscopic}. 

One of the most realistic rough surfaces is modeled using the random process theory \cite{khlyupin2017random, herminghaus2012universal}, where the roughness in the normal direction corresponds to the variance $var$ and the lateral surface structure is defined by the correlation length $cl$. In the works of \cite{khlyupin2017random, aslyamov2017density}, the authors developed a novel approach entitled random surface density functional theory (RS-DFT) that considers rough surfaces as correlated random processes. As demonstrated in \cite{aslyamov2017density, aslyamov2019random}, RS-DFT fits experimental data well and correctly describes fluid properties near rough surfaces. In Figure \ref{fig:rough_surface}, one can find an illustration of rough surfaces obtained from RS-DFT calculation as some realization of a random correlated Gaussian process with certain variance $var$ and correlation length $cl$. A detailed description of RS-DFT can be found in the work of \cite{aslyamov2017density}; here, we present the final equation for the fluid density distribution $\rho(z)$ as a function of the distance to the surface in the normal direction $z$:
 
\begin{widetext}
\begin{equation}
\label{eq:RSDFT_dens}
\rho(z)=\rho_0\exp\left\{-\frac{1}{S(z,var)}\frac{\delta}{\delta \rho}(F_\text{exc}[S,\rho(z)]-F^0_\text{exc}[\rho_0])-U_\text{eff}(z,var,cl)\right\},
\end{equation}
\end{widetext}

\noindent where $\rho_0$ is the bulk fluid density; $F^0_\text{exc}$ and $F_\text{exc}$ are the fluid bulk and the inhomogeneous excess free energies containing the terms of molecular attraction and repulsion, respectively. Equation (\ref{eq:RSDFT_dens}) differs from the popular DFT version \cite{roth2010fundamental} by two modifications disappearing in the limit of an ideally smooth surface only. The first modification is the external effective solid-fluid potential $U_\text{eff}$, which accounts for the surface roughness. A detailed discussion and explicit expression of the effective potential for correlated random surfaces can be found in \cite{khlyupin2017random}. The authors of \cite{khlyupin2017random} developed an averaging procedure based on an analogy with a first passage time probability problem in the theory of Markovian random processes. This approach allows us to describe the molecular interaction with two-parametric rough surfaces using one-dimensional potential $U_\text{eff}(z, var, cl)$ depending on both the variance $var$ and correlation length $cl$.
The second modification is dedicated to proper calculation of the surface area that is available for the fluid molecules at a certain distance $z$. The spatial integration over the lateral plane $dxdy$ takes into account the surface geometry as follows:

\begin{equation}
\label{eq:dxdy}
    \int dz\rho(z)\int dxdy ...=\int dz\rho(z)S(z),
\end{equation}
where $S(z)$ is the surface area free from the solid at the level $z$, which has the following expression:
\begin{equation}
    \label{eq:S_z}
    S(z)=A\left(1-\frac{1}{2} \erfc \frac{z}{\sqrt{2}var}\right).
\end{equation}

\noindent This function depends on the variance $var$ and tends to the total surface area $A$ in the smooth surface limit ($var\to0)$ or with increasing $z$. As a result, the adsorption isotherm for the rough surfaces can be calculated using the solution $\rho(z,var,cl)$ of (\ref{eq:RSDFT_dens}) and the modified surface area $S(z)$:

\begin{equation}
\label{eq:ads_isotherm}
N_\text{ads}=\int dz \rho(z; var, cl)S(z; A, var)-\rho_0 A (H-d),
\end{equation}
where $H$ is the pore width and $d$ is the molecular diameter of the solid-fluid Lennard-Jones interaction. In our work, the solid-fluid interaction parameters are obtained using Lorentz-Berthelot rules with the standard parameters for nitrogen and silica. As one can see from expression (\ref{eq:ads_isotherm}), the number of molecules depends on the roughness parameters $var$, $cl$ and the specific surface area $A$.

The experimental adsorption capacity corresponds to a gram of the solid samples, so $A$ plays the role of the specific surface area measured in $m^2/g$. The surface area $A$ arises from proper integration over the lateral coordinates $dxdy$ (\ref{eq:dxdy}) and does not depend on the surface roughness. In other words, $A$ is the specific surface area of a flat plane that covers the considered rough surface. It is possible to estimate the value range of $A$ using the BET method \cite{thommes2015physisorption} as $A_\text{BET}=n_m a/M$, where $n_m$ is the monolayer capacity, $a$ is the molecular cross-sectional area of the adsorbed molecule in the complete monolayer, and $M$ is the mass of the adsorbent. The surface analysis from the BET method is always limited by the pressure, which ranges from 0.05 to 0.3 in terms of relative pressure $P/P_0$, where $P_0$ is the saturation pressure. The pressure range of the BET model is not the most suitable for surface geometry characterization. The major reason is that an adsorbed film that obscures the surface geometry is influenced by complex confined fluid effects, especially in the case of random rough surfaces \cite{khlyupin2016effects}. The idea to use the surface area as a variable was discussed in the work of \cite{ustinov2005application}, where the author also noted significant sensitivity of $A_\text{BET}$ (deviation in $20\%$) to the pressure range of the experimental data analysis \cite{jaroniec1999standard}. Thus, in our work, we considered $A$ as an additional parameter in the value range defined by the BET method as $0.75 A_\text{BET}\leq A \leq 1.25 A_\text{BET}$.

As one can see from expression (\ref{eq:RSDFT_dens}), the adsorption capacity (\ref{eq:ads_isotherm}) depends on the geometrical parameters even in the case of rare gas molecules near a solid. Unlike the authors of other published approaches, we characterized the surfaces at a region of low pressures $P/P_0<0.1$, where the influence of the solid geometry on the adsorption properties is the highest and does not intersect with the confined fluid effects \cite{ustinov2008nitrogen}.
Thus, in our workflow, experimental nitrogen adsorption isotherms at $77$ K are used as input data. Then, RS-DFT provides the variance $var$ and correlation length $cl$ using the minimization of the following deviation from the experimental data at the certain specific surface area of $0.75 A_\text{BET}\leq A \leq 1.25 A_\text{BET}$:

\begin{equation}
    \label{eq:measure}
   \Delta(A)= \min_{var, cl}\sum_{i=1}^M\left(\frac{N_\text{exp}^{(i)}-N_\text{ads}^{(i)}(var,cl,A)}{N_\text{exp}^{(i)}}\right)^2,
\end{equation}
where the index $i$ corresponds to the relative pressures from the following low pressure range $0.005\leq p_i/p_0\leq 0.1$ and $N_\text{exp}^{i}$ is the experimental data.

\begin{figure}[b]
\includegraphics[width=0.45 \textwidth]{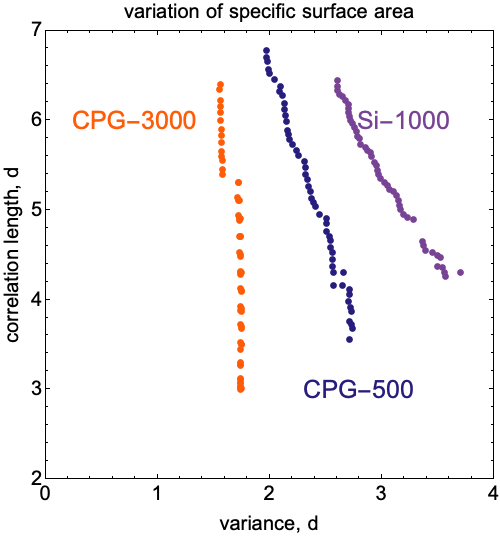}
\caption{\label{fig:S_var-cl} The dots correspond to the best fit surface roughness (the variance and the correlation length) as minimization of (\ref{eq:measure}) for the variation of the specific surface area $A$ in the range $0.75 A_\text{BET}\leq A \leq 1.25 A_\text{BET}$}
\end{figure}

We investigated the surface geometry of popular commercially available meso- and macroporous silica materials. These materials demonstrate a high variety of pore sizes and forms depending on the synthesis method. For example, CPG is synthesized by the Vycor process, which contains phase decomposition due to the melting of the three-component mixture and further dissolving of the less chemically stable phase in the acid. In accordance with patent \cite{haller1973porous}, the pore size of CPG is fully controlled by the temperature and duration of the heat treatment. In addition, the experimental investigations of Vycor glass claim significant roughness around $15-20$ \r{A}. To investigate how CPG nanoroughness depends on the synthesis conditions, we measured low temperature nitrogen adsorption on CPG-500, CPG-1000 and CPG-3000 samples provided by Millipore with pore sizes of approximately $500$ \r{A}, $1000$ \r{A} and $3000$ \r{A}, respectively. As one can see from \cite{haller1973porous}, CPG-3000 was synthesized at much higher temperatures than were CPG-500 and CPG-1000, which should be reflected in the comparison of their surface characterizations. We considered analog of Vycor glass entitled Varapor-100 by Advanced Glass and Ceramics, USA. We obtained adsorption isotherms for all glass samples in our laboratory using the ASAP-2020 instrument by Micromeritics, USA. We also used the published data on nitrogen adsorption on silica gel LiChrospher Si-1000 from \cite{jaroniec1999standard}. Thus, the set of considered materials is complete enough to compare the influences from different synthesis processes and conditions on the surface roughness.

\begin{figure}[t]
\centering
\includegraphics[width=0.45 \textwidth]{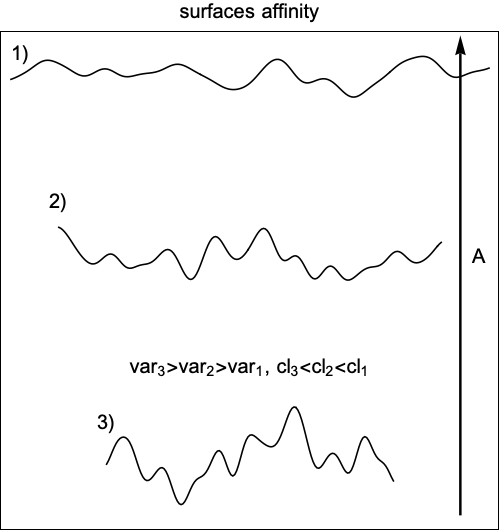}
\caption{The numbered rough lines correspond to the roughness parameters obtained for specific surface area $A$ correspond to 1) $1.25A_\text{BET}$, 2) $A_\text{BET}$ and 3) $0.75A_\text{BET}$. As one can see, to adsorb the same number of molecules on a decreasing surface area, the geometry becomes more rough. In the case of a smaller specific area $0.75A_\text{BET}$, the variance is larger, and thet correlation length is smaller.}
\label{fig:S_variation_sketch}
\end{figure}

\begin{figure*}[t]
\centering
\includegraphics[width=\textwidth]{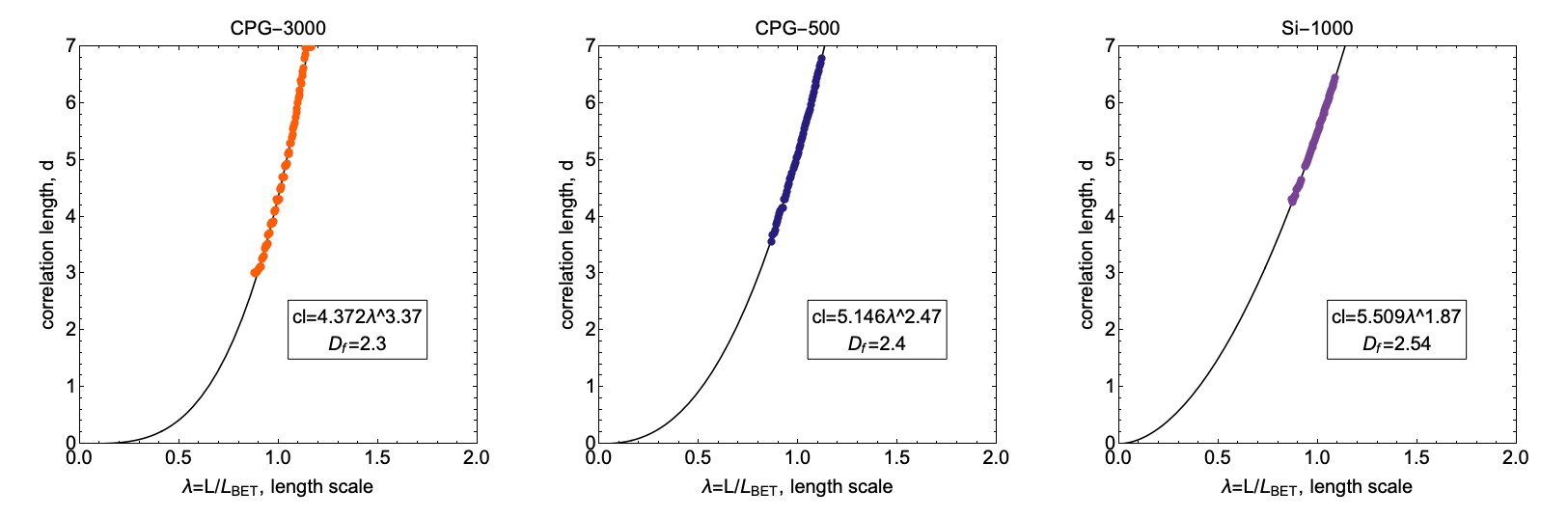}
\caption{ \label{fig:S_L-cl} Correlation length $cl$ corresponding to the minimum (\ref{eq:measure}) as a function of the specific length $L=A^{1/2}$. Accounting the limit properties, the calculated points (color dots) can be fitted by the curve $cl=k L^\gamma$. The fitting parameters and calculations fractal dimensions from (\ref{eq:fractal}) are shown in the insets.}
\end{figure*}

In our study, the experimental data correspond to the low pressure ($P/P_0<0.1$) nitrogen adsorption measurements at 77 K.
We measured the silica glass samples (CPG, Varapor) in our laboratory using an ASAP2020 instrument. Additionally, we used published tabulated data for LiChrospher Si-1000 from the work of \cite{jaroniec1999standard}. The theoretical approach of RS-DFT provides the fluid density distribution (\ref{eq:RSDFT_dens}) near the rough silica surface defined by the roughness parameters $var$ and $cl$. Considering the experimental data for a certain solid sample, one can obtain the best fit roughness parameters using the minimization of (\ref{eq:measure}) over $1.5 d\leq var\leq4d$ and $3 d\leq cl\leq7d$. This result depends on the specific surface area $A$, so we calculated the set of parameters $var$ and $cl$ (\ref{eq:measure}) for the surface area variation in the range of $0.75 A_\text{BET}\leq  A \leq 1.25 A_\text{BET}$ (with step $0.01 A_\text{BET}$). Figure \ref{fig:S_var-cl} contains the results corresponding to three materials that are the most different from each other: CPG-3000, CPG-500, and Si-1000.

As shown in Fig.~\ref{fig:S_var-cl}, the variation in specific surface area $A$ induces the consequence of similar surfaces. Actually, the points of one color in Fig.~\ref{fig:S_var-cl} correspond to the best fit geometry, allowing the adsorption of the fluid with the same (known from the experiment) value on a certain surface area $A$. Therefore, to adsorb the constant number of molecules on the decreasing surface area, more heterogeneous geometry and deeper fluid penetration into the media are needed (Fig.~\ref{fig:S_variation_sketch}). As shown in Fig.~\ref{fig:S_variation_sketch}, when $A$ increases (black arrow direction), the best fit parameters tend to the limit $var\to0$ and $cl\to\infty$, which is the ideal smooth surface. Thus, $A$ can be considered a measure of roughness that leads to the criteria for surface affinity in terms of fractal dimension. 

Our analysis is similar to the method of \cite{farin1985applications}, where authors considered molecules with size $d$ as the scale unit on a fractal line. Let us consider a rough line of length $L=\sqrt{A}$, and the surface fractal dimension $D_f$ is defined via line-dimension as $D_f=D_\text{line}+1$ \cite{farin1985applications}. The number of patterns on length $L$ can be estimated as $N_d=L/d$, and the surface fractal dimension $D_f$ is defined by the following expression:

\begin{equation}
\label{eq:fractal_def}
D_f=-\lim_{d\to 0}\frac{\log N_d}{\log d}+1=-\lim_{d\to 0}\frac{\log L}{\log d}+2.
\end{equation}

To determine how the length $L$ is measured by the molecular diameter $d$, one can consider the dependence of $L$ on the parameters $var$ and $cl$. As the points $var$ and $cl$ for $A=L^2$ variation in Fig.~\ref{fig:S_var-cl} can be fitted well by a one-dimensional curve, it is possible to consider a new pair of parameters, for example, $cl$ and $L$.

\begin{table}[b]
 \caption{\label{tab:fractal} Fractal dimensions $D_f$ calculated from RS-DFT}  
 \begin{ruledtabular}
 \begin{tabular}{cccc}
    Material     & k & $\gamma$     & $D_f$ \\
    \hline
    CPG-500a & 5.046 & 2.47 & 2.4  \\
    CPG-500b & 5.36 & 2.62 & 2.38  \\
    CPG-1000 & 5.638 & 2.42 & 2.41 \\
    CPG-3000  & 4.372 & 3.37 & 2.3 \\  
    Varapor-100 & 5 & 2.44 & 2.41 \\
     Si-1000 & 5.51 & 1.87 & 2.54   \\
  \end{tabular}
  \end{ruledtabular}
\end{table}

Fig.~\ref{fig:S_L-cl} demonstrates how the points from Fig.~\ref{fig:S_var-cl} can be rewritten in terms of the correlation length $cl$ as a function of the length $L$. Taking into account the condition $cl(0)=0$, the correlation length $cl$ can be represented as the power function $cl=k L^\gamma$, where $k>0$ and $\gamma>0$ are the fitting parameters. These parameters can be found in the insets of Fig.~\ref{fig:S_L-cl} and in Table \ref{tab:fractal}. However, for molecules with different sizes near a certain rough surface, one can obtain the proportionality for the correlation length as $cl\sim 1/d$. Now, the fractal dimension can be calculated from (\ref{eq:fractal_def}) using the expression $L\sim d^{-1/\gamma}$ as follows:

\begin{equation}
\label{eq:fractal}
    D_f=2+\frac{1}{\gamma}
\end{equation}

\begin{figure*}[t]
    \centering
    \includegraphics[width=\textwidth]{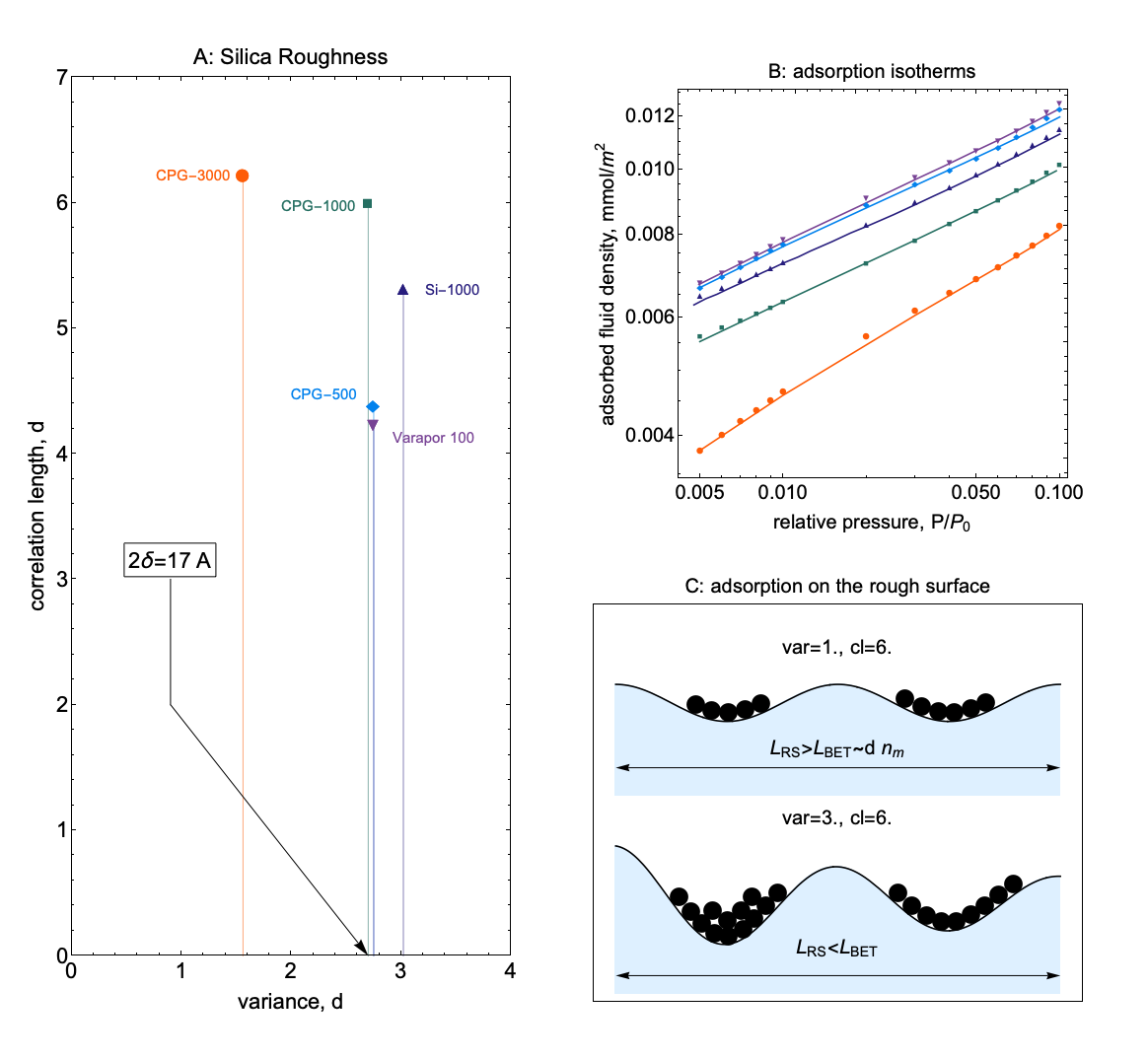}
    \caption{\label{fig:adsorption_roughness}
    \textbf{A}: Best fit roughness parameters $var$ and $cl$ corresponding to the specific surface area $A_\text{RS}$, which provides the minimal deviation (\ref{eq:measure}) among $0.75 A_\text{BET}\leq A \leq 1.25 A_\text{BET}$. The framed inset $2\delta=17 A$ is the roughness region ($2 var$) of the materials with $D_f=2.4$. This result fits experimental SAXS data demonstrating the roughness region in the range of $15\leq2\delta\leq 20$.
    \textbf{B}: The solid lines are interpolated from the experimental measurements of nitrogen adsorption at $77K$ on the silica materials (the colors of the materials are the same as those in Fig.~A). The dots are points used in the minimization of \ref{eq:measure} at $A_\text{RS}$ and roughness parameters from Fig.~A.
\textbf{C}: Sketch of adsorption on rough surfaces illustrating the comparison of $A_\text{RS}$ and $A_\text{BET}$. (Top sketch) Adsorbed molecules on more smooth surfaces (CPG-3000) form mono-layer clusters in small cavities. The number of these molecules defines the BET surface, which is smaller than the full two-dimensional coverage. (Bottom sketch) In the case of significantly rough surfaces, the molecules demonstrate that the lack of layer structure and adsorption capacity significantly depends on the roughness. The BET method assuming monolayer adsorption leads to overestimation of the specific surface area and $A_\text{BET}>A_\text{RS}$.
}
\end{figure*}

As one can see from (\ref{eq:fractal}), the fractal dimension depends on only one parameter, $\gamma$. From our study, the fractal dimension can be calculated using the properties of the correlated random surface. In other words, we demonstrated that the random surface model results in a fractal dimension without assumptions about spatial scale self-similarly or self-affinity.

The expression (\ref{eq:fractal}) allows calculation of the fractal dimension from analysis of the adsorption isotherm. In this work, we calculated $D_f$ for silica glass samples, such as CPG-500, CPG-1000, CPG-3000, Varapor-100, and one example of silica gel, LiChrospher Si-1000. The obtained fractal dimensions are presented in Table~\ref{tab:fractal}. The published experimental measurements of silica glass materials using SAXS and Porod's law demonstrate a fractal dimension of $2.4$ \cite{levitz1991porous,mitropoulos1995characterization,coasne2012atomistic}. Thus, as one can see from Table \ref{tab:fractal}, our obtained results for $D_f$ coincide with the experimental data well.

To obtain the detailed rough surface geometry, the fractal dimension is not enough, and the best fit roughness parameters $var$, $cl$ are needed. For this reason, we define the surface area $A_\text{RS}$ corresponding to the minimum of the deviation (\ref{eq:measure}) among the values obtained by the surface variation over $0.75A_\text{BET} \leq A  \leq  1.25 A_\text{BET}$.

\begin{table}
\caption{\label{tab:roughness} Obtained the surface roughness parameters $var$, $cl$ and the specific surface area $A_\text{RS}$}
  \begin{ruledtabular}
  \begin{tabular}{ccccc}
    Material     & $A_\text{RS}$, $m^2/g$ & $var/d$ & $cl/d$ & $\frac{A_\text{RS}-A_\text{BET}}{A_\text{BET}}$       \\
    \hline  
    Varapor-100 & 79.3 & 2.8 & 4.2 & -0.13 \\
   CPG-500 & 50.0 & 2.8 & 4.4 & -0.11 \\
     Si-1000 &24.2 & 3 & 5.3 & -0.07  \\
     CPG-1000 &27.4 & 2.7 & 6.0 & 0.05 \\
    CPG-3000  &12.5 & 1.6 & 6.2 & 0.23  \\  
  \end{tabular}
  \end{ruledtabular}
\end{table}

As discussed above, $A$ defines the specific surface area of the two-dimensional plane covering of the rough surface. Therefore, the obtained $A_\text{RS}$ can be both smaller and larger than $A_\text{BET}$ in terms of the dependence on a certain surface geometry. These different cases can be illustrated by the sketch in Fig.~\ref{fig:adsorption_roughness}C. As one can see in the upper illustration in Fig.~\ref{fig:adsorption_roughness}C, adsorbed molecules on more smooth surfaces form mono-layer clusters in small cavities. The BET approach assumes a layering structure of the adsorbed fluid and $A_\text{BET}=n_m a/M$, where $n_m$ is the monolayer capacity, $a$ is the molecular cross-sectional area of the adsorbed molecule in the complete monolayer, and $M$ is the mass of the adsorbent \cite{thommes2015physisorption}. For this reason, the number of molecules $n_m$ leads to BET surface $A_\text{BET}<A_\text{RS}$, which is smaller than the full two-dimensional covering $A_\text{RS}$. In the case of significant roughness (bottom sketch), the adsorbed fluid demonstrates the lack of a layer structure. In the BET method's pressure range, the rough surface stores more adsorbed fluid $n_\text{RS}$ than does the ideal smooth surface due to the penetration of fluid molecules into solid media. The BET method assuming monolayer adsorption leads to overestimation of the specific surface area and $A_\text{BET}>A_\text{RS}$. As shown in Table \ref{tab:roughness}, the obtained results are consistent with these models.

The best fit roughness parameters $var$ and $cl$ corresponding to the specific surface area $A_\text{RS}$ are shown in Fig.~\ref{fig:adsorption_roughness}A. A comparison of the theoretical RS-DFT calculations with the experimental data can be found in Fig.~\ref{fig:adsorption_roughness}B. Despite the similar values of the fractal dimension, the detailed surface geometries are notably different. However, as one can see, the roughness region of glass materials with $D_f=2.4$ is approximately $2\delta=17A$, which is consistent with the results of the SAXS experiments $15\leq2\delta\leq20$ and $D_f=2.4$ for silica glass \cite{levitz1991porous}.

The best fit roughness parameters $var$ and $cl$ corresponding to the specific surface area $A_\text{RS}$ are shown in Fig.~\ref{fig:adsorption_roughness}A. The comparison of theoretical RS-DFT calculations with experimental data can be found in Fig.~\ref{fig:adsorption_roughness}B. In spite of the similar values of the fractal dimension detailed surface geometries are notably different. However, as one can see the roughness region of glass materials with $D_f=2.4$ is around $2\delta=17A$ that is consistent with results of SAXS experiments $15\leq2\delta\leq20$ and $D_f=2.4$ for silica glass \cite{levitz1991porous}.

\begin{figure}[t]
    \centering
  \includegraphics[width=0.45 \textwidth]{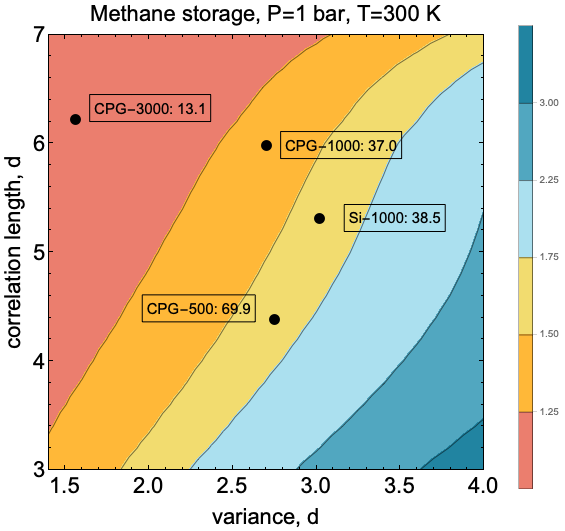}
 \caption{Methane storage on a rough silica surface at atmosphere pressure and room temperature. The contours correspond to constant adsorption storage $N_\text{ads}(var, cl)/N_{0}$. The values of $A_\text{RS} N_\text{ads}/N_{0}$ corresponding to the CPG and Si-1000 samples are presented in the frames and are measured per gram.}
    \label{fig:methane}
\end{figure}

Table \ref{tab:roughness} contains the obtained roughness parameters, the area $A_\text{RS}$ and the relative deviation $\frac{A_\text{RS}-A_\text{BET}}{A_\text{BET}}$. In the work of \cite{ustinov2005equilibrium}, the authors used a specific surface area of LiChrospher Si-1000 as adjustment parameters and found that only a value of 24 $m^2/g$ fits the experimental data. Thus, their result coincides with the obtained specific surface area for Si-1000 $A_\text{RS}=24.15 \; m^2/g$. As discussed above and illustrated in Fig.~\ref{fig:adsorption_roughness}C, our calculations in the case of more smooth surfaces (CPG-3000, CPG-1000) provide the specific surface area $A_\text{RS}>A_\text{BET}$. In the case of significant roughness (CPG-500, Varapor-100, Si-1000) due to enhanced adsorption, the obtained surface coverage is smaller than the BET specific surface $A_\text{RS}<A_\text{BET}$. As shown in Table \ref{tab:roughness}, this adsorption model is reflected in the obtained roughness parameters $var$ and $cl$.

We considered the problem of CH$_4$ storage in mesoporous silica at atmosphere pressure $P=1$ bar and room temperature $T=300$ K. The adsorption capacity $N_\text{ads}(var,cl)/N_0$ accounting for silica surface roughness is shown in Fig.~\ref{fig:methane}, where $N_0$ is the number of adsorbed molecules on the smooth silica surface. The contours in Fig.~\ref{fig:methane} define the isolines in terms of $var$ and $cl$ corresponding to the constant relative adsorption capacity $N_\text{ads}/N_0$. As one can see from the plot legend of Fig.~\ref{fig:methane}, the specific methane storage $N_\text{ads}$ can be notably enhanced by geometrical heterogeneity. To investigate the storage properties of silica samples, we propose a new characteristic:

\begin{equation}
\label{eq:storage}
    C=A_\text{RS}N_\text{ads}(var,cl)/N_0.
\end{equation}

\noindent The relative capacity (\ref{eq:storage}) depends on not only the roughness parameters $var$ and $cl$ but also the specific surface area $A_\text{RS}$ supported the rough geometry. The methane storage capacity $C$ calculated for the CPG and Si-1000 samples using the parameters from Table~\ref{tab:roughness} are shown inside the frames in Fig.~\ref{fig:methane}. The advantage of sample Si-1000 over CPG-1000 due to more rough surface geometry (in terms of both $D_f$ and the roughness parameters) is suppressed by the deviation in the calculated specific surface areas $A_\text{RS}$. Considering the samples with similar roughness, CPG-500 has a clear advantage over Si-1000 because of the notably larger area $A_\text{RS}$. The proposed method of surface geometry characterization in terms of three parameters $A_\text{RS}$, $var$ and $cl$ opens up new opportunities for the search of optimal storage materials considering functional $A_\text{RS}N_\text{ads}(var,cl)$. 

\bibliography{sample}

\begin{thebibliography}{35}%
\makeatletter
\providecommand \@ifxundefined [1]{%
 \@ifx{#1\undefined}
}%
\providecommand \@ifnum [1]{%
 \ifnum #1\expandafter \@firstoftwo
 \else \expandafter \@secondoftwo
 \fi
}%
\providecommand \@ifx [1]{%
 \ifx #1\expandafter \@firstoftwo
 \else \expandafter \@secondoftwo
 \fi
}%
\providecommand \natexlab [1]{#1}%
\providecommand \enquote  [1]{``#1''}%
\providecommand \bibnamefont  [1]{#1}%
\providecommand \bibfnamefont [1]{#1}%
\providecommand \citenamefont [1]{#1}%
\providecommand \href@noop [0]{\@secondoftwo}%
\providecommand \href [0]{\begingroup \@sanitize@url \@href}%
\providecommand \@href[1]{\@@startlink{#1}\@@href}%
\providecommand \@@href[1]{\endgroup#1\@@endlink}%
\providecommand \@sanitize@url [0]{\catcode `\\12\catcode `\$12\catcode
  `\&12\catcode `\#12\catcode `\^12\catcode `\_12\catcode `\%12\relax}%
\providecommand \@@startlink[1]{}%
\providecommand \@@endlink[0]{}%
\providecommand \url  [0]{\begingroup\@sanitize@url \@url }%
\providecommand \@url [1]{\endgroup\@href {#1}{\urlprefix }}%
\providecommand \urlprefix  [0]{URL }%
\providecommand \Eprint [0]{\href }%
\providecommand \doibase [0]{http://dx.doi.org/}%
\providecommand \selectlanguage [0]{\@gobble}%
\providecommand \bibinfo  [0]{\@secondoftwo}%
\providecommand \bibfield  [0]{\@secondoftwo}%
\providecommand \translation [1]{[#1]}%
\providecommand \BibitemOpen [0]{}%
\providecommand \bibitemStop [0]{}%
\providecommand \bibitemNoStop [0]{.\EOS\space}%
\providecommand \EOS [0]{\spacefactor3000\relax}%
\providecommand \BibitemShut  [1]{\csname bibitem#1\endcsname}%
\let\auto@bib@innerbib\@empty
\bibitem [{\citenamefont {F{\'e}rey}\ \emph {et~al.}(2011)\citenamefont
  {F{\'e}rey}, \citenamefont {Serre}, \citenamefont {Devic}, \citenamefont
  {Maurin}, \citenamefont {Jobic}, \citenamefont {Llewellyn}, \citenamefont
  {De~Weireld}, \citenamefont {Vimont}, \citenamefont {Daturi},\ and\
  \citenamefont {Chang}}]{ferey2011hybrid}%
  \BibitemOpen
  \bibfield  {author} {\bibinfo {author} {\bibfnamefont {G.}~\bibnamefont
  {F{\'e}rey}}, \bibinfo {author} {\bibfnamefont {C.}~\bibnamefont {Serre}},
  \bibinfo {author} {\bibfnamefont {T.}~\bibnamefont {Devic}}, \bibinfo
  {author} {\bibfnamefont {G.}~\bibnamefont {Maurin}}, \bibinfo {author}
  {\bibfnamefont {H.}~\bibnamefont {Jobic}}, \bibinfo {author} {\bibfnamefont
  {P.~L.}\ \bibnamefont {Llewellyn}}, \bibinfo {author} {\bibfnamefont
  {G.}~\bibnamefont {De~Weireld}}, \bibinfo {author} {\bibfnamefont
  {A.}~\bibnamefont {Vimont}}, \bibinfo {author} {\bibfnamefont
  {M.}~\bibnamefont {Daturi}}, \ and\ \bibinfo {author} {\bibfnamefont {J.-S.}\
  \bibnamefont {Chang}},\ }\href@noop {} {\bibfield  {journal} {\bibinfo
  {journal} {Chemical Society Reviews}\ }\textbf {\bibinfo {volume} {40}},\
  \bibinfo {pages} {550} (\bibinfo {year} {2011})}\BibitemShut {NoStop}%
\bibitem [{\citenamefont {Hartmann}(2005)}]{hartmann2005ordered}%
  \BibitemOpen
  \bibfield  {author} {\bibinfo {author} {\bibfnamefont {M.}~\bibnamefont
  {Hartmann}},\ }\href@noop {} {\bibfield  {journal} {\bibinfo  {journal}
  {Chemistry of materials}\ }\textbf {\bibinfo {volume} {17}},\ \bibinfo
  {pages} {4577} (\bibinfo {year} {2005})}\BibitemShut {NoStop}%
\bibitem [{\citenamefont {Wei}\ \emph {et~al.}(2018)\citenamefont {Wei},
  \citenamefont {Yan}, \citenamefont {Wang},\ and\ \citenamefont
  {Yang}}]{wei2018fabrication}%
  \BibitemOpen
  \bibfield  {author} {\bibinfo {author} {\bibfnamefont {L.}~\bibnamefont
  {Wei}}, \bibinfo {author} {\bibfnamefont {S.}~\bibnamefont {Yan}}, \bibinfo
  {author} {\bibfnamefont {H.}~\bibnamefont {Wang}}, \ and\ \bibinfo {author}
  {\bibfnamefont {H.}~\bibnamefont {Yang}},\ }\href@noop {} {\bibfield
  {journal} {\bibinfo  {journal} {NPG Asia Materials}\ }\textbf {\bibinfo
  {volume} {10}},\ \bibinfo {pages} {899} (\bibinfo {year} {2018})}\BibitemShut
  {NoStop}%
\bibitem [{\citenamefont {Vilarrasa-Garc{\'\i}a}\ \emph
  {et~al.}(2014)\citenamefont {Vilarrasa-Garc{\'\i}a}, \citenamefont {Cecilia},
  \citenamefont {Santos}, \citenamefont {Cavalcante~Jr}, \citenamefont
  {Jim{\'e}nez-Jim{\'e}nez}, \citenamefont {Azevedo},\ and\ \citenamefont
  {Rodr{\'\i}guez-Castell{\'o}n}}]{vilarrasa2014co2}%
  \BibitemOpen
  \bibfield  {author} {\bibinfo {author} {\bibfnamefont {E.}~\bibnamefont
  {Vilarrasa-Garc{\'\i}a}}, \bibinfo {author} {\bibfnamefont {J.}~\bibnamefont
  {Cecilia}}, \bibinfo {author} {\bibfnamefont {S.}~\bibnamefont {Santos}},
  \bibinfo {author} {\bibfnamefont {C.}~\bibnamefont {Cavalcante~Jr}}, \bibinfo
  {author} {\bibfnamefont {J.}~\bibnamefont {Jim{\'e}nez-Jim{\'e}nez}},
  \bibinfo {author} {\bibfnamefont {D.}~\bibnamefont {Azevedo}}, \ and\
  \bibinfo {author} {\bibfnamefont {E.}~\bibnamefont
  {Rodr{\'\i}guez-Castell{\'o}n}},\ }\href@noop {} {\bibfield  {journal}
  {\bibinfo  {journal} {Microporous and Mesoporous Materials}\ }\textbf
  {\bibinfo {volume} {187}},\ \bibinfo {pages} {125} (\bibinfo {year}
  {2014})}\BibitemShut {NoStop}%
\bibitem [{\citenamefont {Weetall}(1993)}]{weetall1993preparation}%
  \BibitemOpen
  \bibfield  {author} {\bibinfo {author} {\bibfnamefont {H.~H.}\ \bibnamefont
  {Weetall}},\ }\href@noop {} {\bibfield  {journal} {\bibinfo  {journal}
  {Applied biochemistry and biotechnology}\ }\textbf {\bibinfo {volume} {41}},\
  \bibinfo {pages} {157} (\bibinfo {year} {1993})}\BibitemShut {NoStop}%
\bibitem [{\citenamefont {Avnir}\ \emph {et~al.}(1994)\citenamefont {Avnir},
  \citenamefont {Braun}, \citenamefont {Lev},\ and\ \citenamefont
  {Ottolenghi}}]{avnir1994enzymes}%
  \BibitemOpen
  \bibfield  {author} {\bibinfo {author} {\bibfnamefont {D.}~\bibnamefont
  {Avnir}}, \bibinfo {author} {\bibfnamefont {S.}~\bibnamefont {Braun}},
  \bibinfo {author} {\bibfnamefont {O.}~\bibnamefont {Lev}}, \ and\ \bibinfo
  {author} {\bibfnamefont {M.}~\bibnamefont {Ottolenghi}},\ }\href@noop {}
  {\bibfield  {journal} {\bibinfo  {journal} {Chemistry of Materials}\ }\textbf
  {\bibinfo {volume} {6}},\ \bibinfo {pages} {1605} (\bibinfo {year}
  {1994})}\BibitemShut {NoStop}%
\bibitem [{\citenamefont {Levitz}\ \emph {et~al.}(1991)\citenamefont {Levitz},
  \citenamefont {Ehret}, \citenamefont {Sinha},\ and\ \citenamefont
  {Drake}}]{levitz1991porous}%
  \BibitemOpen
  \bibfield  {author} {\bibinfo {author} {\bibfnamefont {P.}~\bibnamefont
  {Levitz}}, \bibinfo {author} {\bibfnamefont {G.}~\bibnamefont {Ehret}},
  \bibinfo {author} {\bibfnamefont {S.}~\bibnamefont {Sinha}}, \ and\ \bibinfo
  {author} {\bibfnamefont {J.}~\bibnamefont {Drake}},\ }\href@noop {}
  {\bibfield  {journal} {\bibinfo  {journal} {The Journal of chemical physics}\
  }\textbf {\bibinfo {volume} {95}},\ \bibinfo {pages} {6151} (\bibinfo {year}
  {1991})}\BibitemShut {NoStop}%
\bibitem [{\citenamefont {Mitropoulos}\ \emph {et~al.}(1995)\citenamefont
  {Mitropoulos}, \citenamefont {Haynes}, \citenamefont {Richardson},\ and\
  \citenamefont {Kanellopoulos}}]{mitropoulos1995characterization}%
  \BibitemOpen
  \bibfield  {author} {\bibinfo {author} {\bibfnamefont {A.~C.}\ \bibnamefont
  {Mitropoulos}}, \bibinfo {author} {\bibfnamefont {J.}~\bibnamefont {Haynes}},
  \bibinfo {author} {\bibfnamefont {R.}~\bibnamefont {Richardson}}, \ and\
  \bibinfo {author} {\bibfnamefont {N.}~\bibnamefont {Kanellopoulos}},\
  }\href@noop {} {\bibfield  {journal} {\bibinfo  {journal} {Physical Review
  B}\ }\textbf {\bibinfo {volume} {52}},\ \bibinfo {pages} {10035} (\bibinfo
  {year} {1995})}\BibitemShut {NoStop}%
\bibitem [{\citenamefont {Mitropoulos}\ \emph {et~al.}(2015)\citenamefont
  {Mitropoulos}, \citenamefont {Stefanopoulos}, \citenamefont {Favvas},
  \citenamefont {Vansant},\ and\ \citenamefont
  {Hankins}}]{mitropoulos2015formation}%
  \BibitemOpen
  \bibfield  {author} {\bibinfo {author} {\bibfnamefont {A.}~\bibnamefont
  {Mitropoulos}}, \bibinfo {author} {\bibfnamefont {K.}~\bibnamefont
  {Stefanopoulos}}, \bibinfo {author} {\bibfnamefont {E.}~\bibnamefont
  {Favvas}}, \bibinfo {author} {\bibfnamefont {E.}~\bibnamefont {Vansant}}, \
  and\ \bibinfo {author} {\bibfnamefont {N.}~\bibnamefont {Hankins}},\
  }\href@noop {} {\bibfield  {journal} {\bibinfo  {journal} {Scientific
  reports}\ }\textbf {\bibinfo {volume} {5}},\ \bibinfo {pages} {10943}
  (\bibinfo {year} {2015})}\BibitemShut {NoStop}%
\bibitem [{\citenamefont {Sonwane}\ \emph {et~al.}(1999)\citenamefont
  {Sonwane}, \citenamefont {Bhatia},\ and\ \citenamefont
  {Calos}}]{sonwane1999characterization}%
  \BibitemOpen
  \bibfield  {author} {\bibinfo {author} {\bibfnamefont {C.}~\bibnamefont
  {Sonwane}}, \bibinfo {author} {\bibfnamefont {S.}~\bibnamefont {Bhatia}}, \
  and\ \bibinfo {author} {\bibfnamefont {N.}~\bibnamefont {Calos}},\
  }\href@noop {} {\bibfield  {journal} {\bibinfo  {journal} {Langmuir}\
  }\textbf {\bibinfo {volume} {15}},\ \bibinfo {pages} {4603} (\bibinfo {year}
  {1999})}\BibitemShut {NoStop}%
\bibitem [{\citenamefont {Smith}\ and\ \citenamefont
  {Lobo}(2010)}]{smith2010fractal}%
  \BibitemOpen
  \bibfield  {author} {\bibinfo {author} {\bibfnamefont {M.~A.}\ \bibnamefont
  {Smith}}\ and\ \bibinfo {author} {\bibfnamefont {R.~F.}\ \bibnamefont
  {Lobo}},\ }\href@noop {} {\bibfield  {journal} {\bibinfo  {journal}
  {Microporous and Mesoporous Materials}\ }\textbf {\bibinfo {volume} {131}},\
  \bibinfo {pages} {204} (\bibinfo {year} {2010})}\BibitemShut {NoStop}%
\bibitem [{\citenamefont {Coasne}\ and\ \citenamefont
  {Ugliengo}(2012)}]{coasne2012atomistic}%
  \BibitemOpen
  \bibfield  {author} {\bibinfo {author} {\bibfnamefont {B.}~\bibnamefont
  {Coasne}}\ and\ \bibinfo {author} {\bibfnamefont {P.}~\bibnamefont
  {Ugliengo}},\ }\href@noop {} {\bibfield  {journal} {\bibinfo  {journal}
  {Langmuir}\ }\textbf {\bibinfo {volume} {28}},\ \bibinfo {pages} {11131}
  (\bibinfo {year} {2012})}\BibitemShut {NoStop}%
\bibitem [{\citenamefont {Ravikovitch}\ and\ \citenamefont
  {Neimark}(2006)}]{ravikovitch2006density}%
  \BibitemOpen
  \bibfield  {author} {\bibinfo {author} {\bibfnamefont {P.~I.}\ \bibnamefont
  {Ravikovitch}}\ and\ \bibinfo {author} {\bibfnamefont {A.~V.}\ \bibnamefont
  {Neimark}},\ }\href@noop {} {\bibfield  {journal} {\bibinfo  {journal}
  {Langmuir}\ }\textbf {\bibinfo {volume} {22}},\ \bibinfo {pages} {11171}
  (\bibinfo {year} {2006})}\BibitemShut {NoStop}%
\bibitem [{\citenamefont {Neimark}\ \emph {et~al.}(2009)\citenamefont
  {Neimark}, \citenamefont {Lin}, \citenamefont {Ravikovitch},\ and\
  \citenamefont {Thommes}}]{neimark2009quenched}%
  \BibitemOpen
  \bibfield  {author} {\bibinfo {author} {\bibfnamefont {A.~V.}\ \bibnamefont
  {Neimark}}, \bibinfo {author} {\bibfnamefont {Y.}~\bibnamefont {Lin}},
  \bibinfo {author} {\bibfnamefont {P.~I.}\ \bibnamefont {Ravikovitch}}, \ and\
  \bibinfo {author} {\bibfnamefont {M.}~\bibnamefont {Thommes}},\ }\href@noop
  {} {\bibfield  {journal} {\bibinfo  {journal} {Carbon}\ }\textbf {\bibinfo
  {volume} {47}},\ \bibinfo {pages} {1617} (\bibinfo {year}
  {2009})}\BibitemShut {NoStop}%
\bibitem [{\citenamefont {Jagiello}\ and\ \citenamefont
  {Olivier}(2013)}]{jagiello2013carbon}%
  \BibitemOpen
  \bibfield  {author} {\bibinfo {author} {\bibfnamefont {J.}~\bibnamefont
  {Jagiello}}\ and\ \bibinfo {author} {\bibfnamefont {J.~P.}\ \bibnamefont
  {Olivier}},\ }\href@noop {} {\bibfield  {journal} {\bibinfo  {journal}
  {Adsorption}\ }\textbf {\bibinfo {volume} {19}},\ \bibinfo {pages} {777}
  (\bibinfo {year} {2013})}\BibitemShut {NoStop}%
\bibitem [{\citenamefont {Jagiello}\ \emph {et~al.}(2015)\citenamefont
  {Jagiello}, \citenamefont {Ania}, \citenamefont {Parra},\ and\ \citenamefont
  {Cook}}]{jagiello2015dual}%
  \BibitemOpen
  \bibfield  {author} {\bibinfo {author} {\bibfnamefont {J.}~\bibnamefont
  {Jagiello}}, \bibinfo {author} {\bibfnamefont {C.}~\bibnamefont {Ania}},
  \bibinfo {author} {\bibfnamefont {J.~B.}\ \bibnamefont {Parra}}, \ and\
  \bibinfo {author} {\bibfnamefont {C.}~\bibnamefont {Cook}},\ }\href@noop {}
  {\bibfield  {journal} {\bibinfo  {journal} {Carbon}\ }\textbf {\bibinfo
  {volume} {91}},\ \bibinfo {pages} {330} (\bibinfo {year} {2015})}\BibitemShut
  {NoStop}%
\bibitem [{\citenamefont {Aslyamov}\ and\ \citenamefont
  {Khlyupin}(2017)}]{aslyamov2017density}%
  \BibitemOpen
  \bibfield  {author} {\bibinfo {author} {\bibfnamefont {T.}~\bibnamefont
  {Aslyamov}}\ and\ \bibinfo {author} {\bibfnamefont {A.}~\bibnamefont
  {Khlyupin}},\ }\href@noop {} {\bibfield  {journal} {\bibinfo  {journal} {The
  Journal of chemical physics}\ }\textbf {\bibinfo {volume} {147}},\ \bibinfo
  {pages} {154703} (\bibinfo {year} {2017})}\BibitemShut {NoStop}%
\bibitem [{\citenamefont {Aslyamov}\ \emph {et~al.}(2019)\citenamefont
  {Aslyamov}, \citenamefont {Pletneva},\ and\ \citenamefont
  {Khlyupin}}]{aslyamov2019random}%
  \BibitemOpen
  \bibfield  {author} {\bibinfo {author} {\bibfnamefont {T.}~\bibnamefont
  {Aslyamov}}, \bibinfo {author} {\bibfnamefont {V.}~\bibnamefont {Pletneva}},
  \ and\ \bibinfo {author} {\bibfnamefont {A.}~\bibnamefont {Khlyupin}},\
  }\href@noop {} {\bibfield  {journal} {\bibinfo  {journal} {The Journal of
  chemical physics}\ }\textbf {\bibinfo {volume} {150}},\ \bibinfo {pages}
  {054703} (\bibinfo {year} {2019})}\BibitemShut {NoStop}%
\bibitem [{\citenamefont {Serna-Guerrero}\ \emph {et~al.}(2010)\citenamefont
  {Serna-Guerrero}, \citenamefont {Belmabkhout},\ and\ \citenamefont
  {Sayari}}]{serna2010influence}%
  \BibitemOpen
  \bibfield  {author} {\bibinfo {author} {\bibfnamefont {R.}~\bibnamefont
  {Serna-Guerrero}}, \bibinfo {author} {\bibfnamefont {Y.}~\bibnamefont
  {Belmabkhout}}, \ and\ \bibinfo {author} {\bibfnamefont {A.}~\bibnamefont
  {Sayari}},\ }\href@noop {} {\bibfield  {journal} {\bibinfo  {journal}
  {Chemical Engineering Science}\ }\textbf {\bibinfo {volume} {65}},\ \bibinfo
  {pages} {4166} (\bibinfo {year} {2010})}\BibitemShut {NoStop}%
\bibitem [{\citenamefont {Chen}\ \emph {et~al.}(2017)\citenamefont {Chen},
  \citenamefont {Zhang}, \citenamefont {Row},\ and\ \citenamefont
  {Ahn}}]{chen2017amine}%
  \BibitemOpen
  \bibfield  {author} {\bibinfo {author} {\bibfnamefont {C.}~\bibnamefont
  {Chen}}, \bibinfo {author} {\bibfnamefont {S.}~\bibnamefont {Zhang}},
  \bibinfo {author} {\bibfnamefont {K.~H.}\ \bibnamefont {Row}}, \ and\
  \bibinfo {author} {\bibfnamefont {W.-S.}\ \bibnamefont {Ahn}},\ }\href@noop
  {} {\bibfield  {journal} {\bibinfo  {journal} {Journal of energy chemistry}\
  }\textbf {\bibinfo {volume} {26}},\ \bibinfo {pages} {868} (\bibinfo {year}
  {2017})}\BibitemShut {NoStop}%
\bibitem [{\citenamefont {Galarneau}\ \emph {et~al.}(2009)\citenamefont
  {Galarneau}, \citenamefont {Calin}, \citenamefont {Iapichella}, \citenamefont
  {Barrande}, \citenamefont {Denoyel}, \citenamefont {Coasne},\ and\
  \citenamefont {Fajula}}]{galarneau2009optimization}%
  \BibitemOpen
  \bibfield  {author} {\bibinfo {author} {\bibfnamefont {A.}~\bibnamefont
  {Galarneau}}, \bibinfo {author} {\bibfnamefont {N.}~\bibnamefont {Calin}},
  \bibinfo {author} {\bibfnamefont {J.}~\bibnamefont {Iapichella}}, \bibinfo
  {author} {\bibfnamefont {M.}~\bibnamefont {Barrande}}, \bibinfo {author}
  {\bibfnamefont {R.}~\bibnamefont {Denoyel}}, \bibinfo {author} {\bibfnamefont
  {B.}~\bibnamefont {Coasne}}, \ and\ \bibinfo {author} {\bibfnamefont
  {F.}~\bibnamefont {Fajula}},\ }\href@noop {} {\bibfield  {journal} {\bibinfo
  {journal} {Chemistry of Materials}\ }\textbf {\bibinfo {volume} {21}},\
  \bibinfo {pages} {1884} (\bibinfo {year} {2009})}\BibitemShut {NoStop}%
\bibitem [{\citenamefont {Smith}\ \emph {et~al.}(2014)\citenamefont {Smith},
  \citenamefont {Zoelle}, \citenamefont {Yang}, \citenamefont {Rioux},
  \citenamefont {Hamilton}, \citenamefont {Amakawa}, \citenamefont {Nielsen},\
  and\ \citenamefont {Trunschke}}]{smith2014surface}%
  \BibitemOpen
  \bibfield  {author} {\bibinfo {author} {\bibfnamefont {M.~A.}\ \bibnamefont
  {Smith}}, \bibinfo {author} {\bibfnamefont {A.}~\bibnamefont {Zoelle}},
  \bibinfo {author} {\bibfnamefont {Y.}~\bibnamefont {Yang}}, \bibinfo {author}
  {\bibfnamefont {R.~M.}\ \bibnamefont {Rioux}}, \bibinfo {author}
  {\bibfnamefont {N.~G.}\ \bibnamefont {Hamilton}}, \bibinfo {author}
  {\bibfnamefont {K.}~\bibnamefont {Amakawa}}, \bibinfo {author} {\bibfnamefont
  {P.~K.}\ \bibnamefont {Nielsen}}, \ and\ \bibinfo {author} {\bibfnamefont
  {A.}~\bibnamefont {Trunschke}},\ }\href@noop {} {\bibfield  {journal}
  {\bibinfo  {journal} {Journal of catalysis}\ }\textbf {\bibinfo {volume}
  {312}},\ \bibinfo {pages} {170} (\bibinfo {year} {2014})}\BibitemShut
  {NoStop}%
\bibitem [{\citenamefont {Shi}\ \emph {et~al.}(2019)\citenamefont {Shi},
  \citenamefont {Santiso},\ and\ \citenamefont {Gubbins}}]{shi2019bottom}%
  \BibitemOpen
  \bibfield  {author} {\bibinfo {author} {\bibfnamefont {K.}~\bibnamefont
  {Shi}}, \bibinfo {author} {\bibfnamefont {E.~E.}\ \bibnamefont {Santiso}}, \
  and\ \bibinfo {author} {\bibfnamefont {K.~E.}\ \bibnamefont {Gubbins}},\
  }\href@noop {} {\bibfield  {journal} {\bibinfo  {journal} {Langmuir}\
  }\textbf {\bibinfo {volume} {35}},\ \bibinfo {pages} {5975} (\bibinfo {year}
  {2019})}\BibitemShut {NoStop}%
\bibitem [{\citenamefont {Carr}\ \emph {et~al.}(2011)\citenamefont {Carr},
  \citenamefont {Comer}, \citenamefont {Ginsberg},\ and\ \citenamefont
  {Aksimentiev}}]{carr2011microscopic}%
  \BibitemOpen
  \bibfield  {author} {\bibinfo {author} {\bibfnamefont {R.}~\bibnamefont
  {Carr}}, \bibinfo {author} {\bibfnamefont {J.}~\bibnamefont {Comer}},
  \bibinfo {author} {\bibfnamefont {M.~D.}\ \bibnamefont {Ginsberg}}, \ and\
  \bibinfo {author} {\bibfnamefont {A.}~\bibnamefont {Aksimentiev}},\
  }\href@noop {} {\bibfield  {journal} {\bibinfo  {journal} {The journal of
  physical chemistry letters}\ }\textbf {\bibinfo {volume} {2}},\ \bibinfo
  {pages} {1804} (\bibinfo {year} {2011})}\BibitemShut {NoStop}%
\bibitem [{\citenamefont {Khlyupin}\ and\ \citenamefont
  {Aslyamov}(2017)}]{khlyupin2017random}%
  \BibitemOpen
  \bibfield  {author} {\bibinfo {author} {\bibfnamefont {A.}~\bibnamefont
  {Khlyupin}}\ and\ \bibinfo {author} {\bibfnamefont {T.}~\bibnamefont
  {Aslyamov}},\ }\href@noop {} {\bibfield  {journal} {\bibinfo  {journal}
  {Journal of Statistical Physics}\ }\textbf {\bibinfo {volume} {167}},\
  \bibinfo {pages} {1519} (\bibinfo {year} {2017})}\BibitemShut {NoStop}%
\bibitem [{\citenamefont {Herminghaus}(2012)}]{herminghaus2012universal}%
  \BibitemOpen
  \bibfield  {author} {\bibinfo {author} {\bibfnamefont {S.}~\bibnamefont
  {Herminghaus}},\ }\href@noop {} {\bibfield  {journal} {\bibinfo  {journal}
  {Physical review letters}\ }\textbf {\bibinfo {volume} {109}},\ \bibinfo
  {pages} {236102} (\bibinfo {year} {2012})}\BibitemShut {NoStop}%
\bibitem [{\citenamefont {Roth}(2010)}]{roth2010fundamental}%
  \BibitemOpen
  \bibfield  {author} {\bibinfo {author} {\bibfnamefont {R.}~\bibnamefont
  {Roth}},\ }\href@noop {} {\bibfield  {journal} {\bibinfo  {journal} {Journal
  of Physics: Condensed Matter}\ }\textbf {\bibinfo {volume} {22}},\ \bibinfo
  {pages} {063102} (\bibinfo {year} {2010})}\BibitemShut {NoStop}%
\bibitem [{\citenamefont {Thommes}\ \emph {et~al.}(2015)\citenamefont
  {Thommes}, \citenamefont {Kaneko}, \citenamefont {Neimark}, \citenamefont
  {Olivier}, \citenamefont {Rodriguez-Reinoso}, \citenamefont {Rouquerol},\
  and\ \citenamefont {Sing}}]{thommes2015physisorption}%
  \BibitemOpen
  \bibfield  {author} {\bibinfo {author} {\bibfnamefont {M.}~\bibnamefont
  {Thommes}}, \bibinfo {author} {\bibfnamefont {K.}~\bibnamefont {Kaneko}},
  \bibinfo {author} {\bibfnamefont {A.~V.}\ \bibnamefont {Neimark}}, \bibinfo
  {author} {\bibfnamefont {J.~P.}\ \bibnamefont {Olivier}}, \bibinfo {author}
  {\bibfnamefont {F.}~\bibnamefont {Rodriguez-Reinoso}}, \bibinfo {author}
  {\bibfnamefont {J.}~\bibnamefont {Rouquerol}}, \ and\ \bibinfo {author}
  {\bibfnamefont {K.~S.}\ \bibnamefont {Sing}},\ }\href@noop {} {\bibfield
  {journal} {\bibinfo  {journal} {Pure and Applied Chemistry}\ }\textbf
  {\bibinfo {volume} {87}},\ \bibinfo {pages} {1051} (\bibinfo {year}
  {2015})}\BibitemShut {NoStop}%
\bibitem [{\citenamefont {Khlyupin}(2016)}]{khlyupin2016effects}%
  \BibitemOpen
  \bibfield  {author} {\bibinfo {author} {\bibfnamefont {A.}~\bibnamefont
  {Khlyupin}},\ }\href@noop {} {\bibfield  {journal} {\bibinfo  {journal}
  {Journal of Physics: Conference Series}\ }\textbf {\bibinfo {volume} {774}},\
  \bibinfo {pages} {012024} (\bibinfo {year} {2016})}\BibitemShut {NoStop}%
\bibitem [{\citenamefont {Ustinov}\ \emph
  {et~al.}(2005{\natexlab{a}})\citenamefont {Ustinov}, \citenamefont {Do},\
  and\ \citenamefont {Jaroniec}}]{ustinov2005application}%
  \BibitemOpen
  \bibfield  {author} {\bibinfo {author} {\bibfnamefont {E.}~\bibnamefont
  {Ustinov}}, \bibinfo {author} {\bibfnamefont {D.}~\bibnamefont {Do}}, \ and\
  \bibinfo {author} {\bibfnamefont {M.}~\bibnamefont {Jaroniec}},\ }\href@noop
  {} {\bibfield  {journal} {\bibinfo  {journal} {Applied surface science}\
  }\textbf {\bibinfo {volume} {252}},\ \bibinfo {pages} {548} (\bibinfo {year}
  {2005}{\natexlab{a}})}\BibitemShut {NoStop}%
\bibitem [{\citenamefont {Jaroniec}\ \emph {et~al.}(1999)\citenamefont
  {Jaroniec}, \citenamefont {Kruk},\ and\ \citenamefont
  {Olivier}}]{jaroniec1999standard}%
  \BibitemOpen
  \bibfield  {author} {\bibinfo {author} {\bibfnamefont {M.}~\bibnamefont
  {Jaroniec}}, \bibinfo {author} {\bibfnamefont {M.}~\bibnamefont {Kruk}}, \
  and\ \bibinfo {author} {\bibfnamefont {J.~P.}\ \bibnamefont {Olivier}},\
  }\href@noop {} {\bibfield  {journal} {\bibinfo  {journal} {Langmuir}\
  }\textbf {\bibinfo {volume} {15}},\ \bibinfo {pages} {5410} (\bibinfo {year}
  {1999})}\BibitemShut {NoStop}%
\bibitem [{\citenamefont {Ustinov}(2008)}]{ustinov2008nitrogen}%
  \BibitemOpen
  \bibfield  {author} {\bibinfo {author} {\bibfnamefont {E.~A.}\ \bibnamefont
  {Ustinov}},\ }\href@noop {} {\bibfield  {journal} {\bibinfo  {journal}
  {Langmuir}\ }\textbf {\bibinfo {volume} {24}},\ \bibinfo {pages} {6668}
  (\bibinfo {year} {2008})}\BibitemShut {NoStop}%
\bibitem [{\citenamefont {Haller}(1973)}]{haller1973porous}%
  \BibitemOpen
  \bibfield  {author} {\bibinfo {author} {\bibfnamefont {W.}~\bibnamefont
  {Haller}},\ }\href@noop {} {\enquote {\bibinfo {title} {Porous material and
  method of making the same},}\ } (\bibinfo {year} {1973}),\ \bibinfo {note}
  {uS Patent 3,758,284}\BibitemShut {NoStop}%
\bibitem [{\citenamefont {Farin}\ \emph {et~al.}(1985)\citenamefont {Farin},
  \citenamefont {Peleg}, \citenamefont {Yavin},\ and\ \citenamefont
  {Avnir}}]{farin1985applications}%
  \BibitemOpen
  \bibfield  {author} {\bibinfo {author} {\bibfnamefont {D.}~\bibnamefont
  {Farin}}, \bibinfo {author} {\bibfnamefont {S.}~\bibnamefont {Peleg}},
  \bibinfo {author} {\bibfnamefont {D.}~\bibnamefont {Yavin}}, \ and\ \bibinfo
  {author} {\bibfnamefont {D.}~\bibnamefont {Avnir}},\ }\href@noop {}
  {\bibfield  {journal} {\bibinfo  {journal} {Langmuir}\ }\textbf {\bibinfo
  {volume} {1}},\ \bibinfo {pages} {399} (\bibinfo {year} {1985})}\BibitemShut
  {NoStop}%
\bibitem [{\citenamefont {Ustinov}\ \emph
  {et~al.}(2005{\natexlab{b}})\citenamefont {Ustinov}, \citenamefont {Do},\
  and\ \citenamefont {Jaroniec}}]{ustinov2005equilibrium}%
  \BibitemOpen
  \bibfield  {author} {\bibinfo {author} {\bibfnamefont {E.}~\bibnamefont
  {Ustinov}}, \bibinfo {author} {\bibfnamefont {D.}~\bibnamefont {Do}}, \ and\
  \bibinfo {author} {\bibfnamefont {M.}~\bibnamefont {Jaroniec}},\ }\href@noop
  {} {\bibfield  {journal} {\bibinfo  {journal} {The Journal of Physical
  Chemistry B}\ }\textbf {\bibinfo {volume} {109}},\ \bibinfo {pages} {1947}
  (\bibinfo {year} {2005}{\natexlab{b}})}\BibitemShut {NoStop}%
\end{thebibliography}%

\end{document}